\documentclass[twocolumn,showpacs,preprintnumbers,superscriptaddress,amsmath,amssymb]{revtex4}

\usepackage{epsfig}
\usepackage{graphicx}
\usepackage{dcolumn}
\usepackage{bm}
\usepackage{times}
\usepackage{xcolor}
\usepackage[percent]{overpic}
\begin{document}

\title{Comprehensive routing strategy on multilayer networks}
\author{Lei Gao}
\affiliation{Web Sciences Center, University of Electronic
Science and Technology of China, Chengdu 610054, China}
\affiliation{Big data research center, University of Electronic Science and Technology of China, Chengdu 610054, China}

\author{Panpan Shu}
\affiliation{School of Sciences, Xi'an University of Technology, Xi'an
  710054, China}

\author{Ming Tang\footnote{Correspondence to: tangminghan007@gmail.com}}
\affiliation{Web Sciences Center, University of Electronic
Science and Technology of China, Chengdu 610054, China}
\affiliation{Big data research center, University of Electronic Science and Technology of China, Chengdu 610054, China}
\affiliation{School of Information Science Technology, East China Normal University, Shanghai 200241, China}

\author{Wei Wang\footnote{Correspondence to: wwzqbx@hotmail.com}}
\affiliation{Web Sciences Center, University of Electronic
Science and Technology of China, Chengdu 610054, China}
\affiliation{Big data research center, University of Electronic Science and Technology of China, Chengdu 610054, China}

\author{Hui Gao}
\affiliation{Web Sciences Center, University of Electronic
Science and Technology of China, Chengdu 610054, China}
\affiliation{Big data research center, University of Electronic Science and Technology of China, Chengdu 610054, China}

\date{\today}

\begin{abstract}
Designing an efficient routing strategy is of great importance to alleviate traffic congestion in multilayer networks. In this work, we design an effective routing strategy for
multilayer networks by comprehensively considering the roles of nodes' local
structures in micro-level, as well as the macro-level differences in
transmission speeds between different layers. Both numerical
and analytical results indicate that our proposed
routing strategy can reasonably redistribute the traffic load of the low speed
layer to the high speed layer, and thus the traffic capacity of multilayer networks are significantly enhanced compared with the monolayer
low speed networks. There is an optimal combination
of macro- and micro-level control parameters at which can remarkably alleviate the
congestion and thus maximize the
traffic capacity for a given multilayer network. Moreover, we find that increasing
the size and the average degree of the high speed layer can enhance the traffic
capacity of multilayer networks more effectively. We finally verify
that real-world network topology does not invalidate the results.
The theoretical predictions agree well with the numerical simulations.

\end{abstract}


\maketitle
\textbf{Alleviating the congestion in transportation and
communication systems is vital to modern society. For
the purpose of redistributing the traffic load in a low
speed transportation system such as bus net, we can establish
a high speed system (e.g., subway network) in the busy
regions or between the stations with high traffic flow,
and the two systems make up a new multilayer system (i.e.,
multilayer network). Recent years, some investigations
about traffic congestion on multilayer networks were performed,
which mainly focused on the different
roles of layers in a macroscopic level (e.g., different transmission speeds),
or the local structures of nodes within the same layer in a microscopic level,
without taking them into consideration comprehensively.
To this end, we propose a comprehensive
routing strategy on multilayer networks composed of a low and a high speed
network. We introduce a macro- and a micro-level parameter to
adjust the roles of network structures played in the routing strategy. Our
strategy redistributes the traffic load in
low speed layer to the high speed layer reasonably,
and the traffic capacity of multilayer networks
are thus remarkably enhanced compared with the monolayer low
speed networks. For a given
multilayer network, an optimal
combination of macro- and micro-level parameters is found.
Under these parameters, the traffic capacity of the system
reaches its maximum value. Moreover, increasing the networks
size and the average degree of the high speed layer
can enhance the traffic capacity of multilayer networks more effectively.
To quantificationally understand the proposed routing
strategy, we developed a theoretical approach and a remarkable
agreement with numerics is observed in both artificial and
real-world networks. Our research
may stimulate future studies on designing realistic transportation
and communication multilayer networks.}

\section{Introduction} \label{sec:intro}
Many systems in modern society can be described by complex networks,
such as power grids, transportation networks and
social networks~\cite{strogatz2001exploring,albert2002statistical,
boccaletti2006complex,newman2010networks,barabasi2016network}.
Routing on such networked systems to enhance traffic
capacity is a significant issue, and has been widely studied
from the perspective of complex network framework over the past
decades~\cite{arenas2001communication,zhao2005onset,
wu2006transport,de2009congestion,barthelemy2011spatial,wu2013analysis}.
Most studies about routing are focused on monolayer networks.
The studies have revealed that traffic congestion is highly
related to the structures of networks~\cite{boccaletti2006complex,
guimera2002dynamical,guimera2002optimal}. Generally,
there are three widely used techniques to enhance the
throughput of the whole network: (1) modification of
network structures~\cite{danila2006optimal,
danila2007transport,liu2007method,zhang2007enhancing},
(2) optimization of traffic resources allocations~\cite{xia2010optimal,
xiang2013traffic,zhang2011enhancing}, and (3) designing better routing strategies~\cite{yan2006efficient,de2009congestion,tang2009self,
tadic2009jamming,ling2010global,tang2011efficient,
rachadi2013self,gan2013optimal}. Compare
with the first two methods, proposing effective routing
strategies seems to be more practical and thus has attracted
much interest. Among numerous different kinds of
proposed routing strategies, an efficient routing strategy
proposed by Yan and his colleagues is
widely acknowledged for its simplicity and
efficiency~\cite{yan2006efficient}. The strategy
redistributes traffic load in central nodes
to other noncentral nodes and improves the network
throughput significantly. Echenique \emph{et al}.
proposed a novel traffic awareness protocol (TAP) by
considering the waiting time of packets,
in which a node forwards a packet to its neighboring
node according to the shortest effective distance~\cite{echenique2004improved,
echenique2005dynamics}. Some
scholars also proposed strategies for systems with
limited band width~\cite{wu2008traffic,wang2009abrupt}.

With the availability of big data, scholars found that modern
infrastructures are actually significantly interact with
and/or depend on each other, which can be described as multilayer (multiplex) networks~\cite{gao2012networks,kivela2014multilayer,lee2015towards,
boccaletti2014structure}. For example, to redistribute the traffic load
in a low speed transportation network, we can build a new high speed
network in the busy regions or between the high flow stations,
and the two monolayer networks constitute a multilayer network.
Researchers have demonstrated that
the dynamics of~\cite{boccaletti2014structure} and
on~\cite{buldyrev2010catastrophic,wang2014asymmetrically,
gomez2012evolution,aguirre2014synchronization} multilayer
networks are markedly different from monolayer networks.
Until very recently, some researchers studied the traffic
dynamics on multilayer complex networks, i.e., how to
alleviate the traffic congestion in order to enhance the
multilayer network capacity~\cite{morris2012transport,
zhou2013efficient,yagan2013conjoining,
tan2014traffic,sole2016congestion,li2016transportation}.
Interestingly, Sol\'{e}-Ribalta \emph{et al.}~\cite{sole2016congestion}
developed a standardized model of transportation in multilayer
networks, and showed that the structure of multiplex networks
can induce congestion on account of the unbalance of shortest
paths between layers. Morris and Barth\'{e}lemy~\cite{morris2012transport}
analyzed a multilayer network consists of two layers, and
showed that it is possible to obtain an optimal communication
multiplex by balancing the effects between decreasing the
average distance and congestion on a very small subset of edges.

The structures of multilayer networks bring new challenges
when we propose effective routing strategies,
and the task is markedly different from monolayer networks. On one
hand, multilayer networks can relieve the traffic congestion
of the low speed layer by using the high speed layer, however congestion
may be induced in the high speed layer~\cite{brummitt2012suppressing}. Although establishing
high speed transportation networks can improve the traffic capacity of low
speed network, how to reasonably the redistribute traffic loads
is an essential issue.
On the other hand, when designing effective routing strategies we should
(1) take the intra-layer structures into considerations
from microscopic perspective, and (2) consider the
efficiencies of different layers from a macroscopic view.
Previous investigations about traffic congestion on multilayer
networks mainly focused on the macroscopic differences between
layers~\cite{morris2012transport},
or the local structure of nodes within the same layer in a microscopic level~\cite{tan2014traffic}, without taking both of them into
consideration comprehensively.
In this work, we propose an comprehensive routing
strategy on multilayer networks
by incorporating the macroscopic difference of speed between layers
and microscopic distinctions among different nodes in the same layer.
We find that our routing strategy
can redistribute the traffic load in low speed layer to high
speed layer reasonably, and the traffic capacity of multilayer networks
are remarkably enhanced compared with the monolayer low speed networks.
For a given multilayer network, there is an optimal combination
of macro-level parameter and micro-level parameter that
maximize the traffic capacity. Increasing the size and average degree of
the high speed layer can enhance the traffic capacity of
multilayer networks more effectively. Numerical results on artificial
multilayer networks as well as the real Work-Facebook multilayer
network agree well with our analysis.

The outline of the paper is as follows. In Sec. II, we give
a detailed description of our
routing strategy on multilayer networks. In Sec. III, we
suggest theoretical analysis. In Sec. IV, we present our simulation results.
Section V summarizes our results and conclusions.

\section{Model}
\subsection{Network model}
\begin{figure}[htbp]
\centerline{\includegraphics[width=1\linewidth]{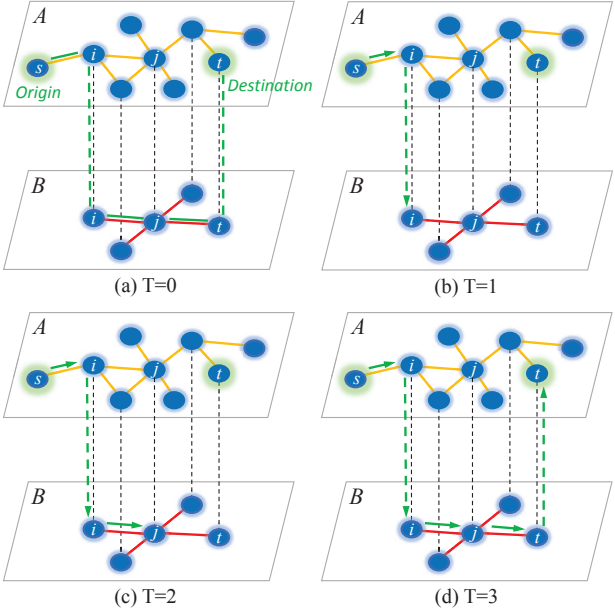}}
\caption{An illustration of multilayer network where the
nodes of layer $B$ form a subset of the nodes of network $A$.
Edges of layer $A$ are shown in orange, edges of network $B$
are shown in red, and nodes in common to both layers are considered to be coupled nodes (shown by dashed lines). Highlighted in green, a path between the `origin' $s$ and the `destination' $t$ is represented, and the arrows show a packet delivery process on the path without congestion. Coupled nodes can deliver packets between layers with no time consumption. (a) At $T=0$, a packet is generated with the randomly chosen`origin' $s$ and `destination' $t$ in layer $A$, and a path (highlighted in green) from node $s$ to $t$ is chosen according to the pre-given routing strategy. (b) $T=1$, `origin' $s$ delivers the packet to the next stop $i$ through the edge ($s,i$) in layer $A$. (c) $T=2$, node $i$ delivers the packet to the next stop $j$ through the edge ($i,j$) in layer $B$. (d) $T=3$, node $j$ delivers the packet to its `destination' $t$ through the edge ($j,t$) in layer $B$, and the packet is removed from the system. }\label{fig1}
\end{figure}

The multilayer network considered is composed by two layers
with $N_A$ and $N_B$ nodes respectively.
Layer $A$ represent the low speed network,
and layer $B$ is the high speed network. In general
condition, the expense of building a high speed network
is far more than that of low speed network, thus the
size of high speed network is smaller. For example, in
the railway-airline multilayer network, where the speed
(cost) of the airline network is faster (more) than that
of the railway network. Thus, the size of the airline
network is smaller than the railway network, and all airline
stations are located at points which can be considered as
nodes in the railway network, but no vice versa~\cite{gu2011onset}.
For simplicity, we assume that the nodes in the high speed layer $B$ are
a random subset of the low speed layer $A$~\cite{morris2012transport}. We use the uncorrelated configuration model
(UCM)~\cite{catanzaro2005generation} to generate the
low speed layer $A$, and use the Erd\"{o}-R\'{e}nyi (ER)
networks~\cite{erd6s1960evolution} to represent the
high speed layer $B$. The multilayer network is generated as
follows: (1) Build layer $A$ using the UCM method with
power-law degree distributions $P(k)\sim k^{-\gamma}$,
where $\gamma$ is the degree exponent. We set the size
of layer $A$ as $N_A$, the minimum degree is
$k_{\rm min}=2$, and the maximum degree is $k_{\rm max}\sim \sqrt{N_A}$.
(2) Randomly select $N_B$ ($N_B\leqslant N_A$) nodes in layer $A$,
and match these nodes one-to-one. This means that each pair of the
two matched nodes $v_A^o$ and $v_B^o$ are actually the same node but in two different transport manners. Both of them can be denoted as a coupled node $v^o$.
Or say, a coupled node $v^o$ has two replica nodes in layers $A$ and $B$, which are denoted by  $v_A^o$ and $v_B^o$ respectively. (3) Construct a ER
network as the second layer $B$ by using the selected $N_B$ nodes in step (2),
i.e., each pair of these randomly selected nodes are connected
with a probability $p$. According to the above three
steps, a multilayer network can be built. Note that
every node in layer $B$ has its counterpart node in layer
$A$, but the inverse is not true. We denote the degree
distribution of the multilayer network as $P(\overrightarrow{K})
=P(k_A,k_B)$, where $k_A$ and $k_B$ denote the degrees
in layer $A$ and $B$ respectively. For a node $v_A$ in layer
$A$ without counterpart in layer $B$, we have $k_B=0$. An illustration
of the multilayer network is shown in Fig.~\ref{fig1}(a).

\subsection{Routing model}
In our model, we assume all nodes in layer $A$ are treated
as both hosts and routers for generating and delivering
packets, while the nodes in layer $B$ can only deliver packets.
We assume that a coupled node $v^o$ can deliver packets between
layers with infinity bandwidth and no time consumption through
two replica nodes $v_A^o$ and $v_B^o$. For simplicity, each node
in the layer $A$ ($B$) has the same maximum packet delivery
ability $C_A$ ($C_B$). That's to say, at each time step
each node $v_A$ can delivery $C_A$ packets to its neighbors in
layer $A$ if it has no counterpart in layer $B$. Otherwise, its
counterpart node $v_B$ can also transmit $C_B$ packets in layer $B$. We
set $C_A= C_B= C=1$ in this paper for simplicity.

Due to the finite delivery ability of nodes, a queue of buffers
is needed for each node to accommodate packets waiting
for being delivered. The transport processes is as follows:
\begin{enumerate}
  \item Packet generation. At each time step, $R$ number
   of packets are generated with randomly chosen origins
   and destinations in layer $A$.
   For each packet, a path from the source to the destination
   is chosen according to the comprehensive multilayer routing
   strategy (to be introduced in the next subsection). If
   there are several paths between these two nodes, we choose one
   randomly. Each newly created packet
   is placed at the end of the queue of its source node $v_A$ if the
   next stop is in layer $A$, or queued at the counterpart node
   $v_B^o$ if the next stop is in layer $B$.

  \item Packet processing. The first-in-first-out
  (FIFO) rule is adopted to hand each queue. At each time step, node $v_A$ ($v_B$) can
  process $C_A=1$ ($C_B=1$) packet from the head of it's queue and deliver the packet
  to the next stop in layer $A$ ($B$). So a coupled node $v^o$
  at most process 2 packets per time step through two replica
  nodes $v_A^o$ and $v_B^o$ in layers $A$ and $B$. For a
  non-coupled node $v_A$ (i.e., node $v_A$ has no counterpart
  in layer $B$), it can process only $C_A=1$ packet.
  When a packet arrives at its destination, it is removed
  from the system; otherwise it is queued.
\end{enumerate}

\subsection{Comprehensive multilayer routing strategy}
By integrating different roles of nodes in micro-level, as
well as different transmission speed of layers in macro-level,
we propose a comprehensive multilayer routing (CMR) strategy,
which can remarkably enhance the traffic capacity
of multilayer networks. We denote that a path between nodes $s$ and $t$ as
\begin{equation}\label{eq1}
p(s \rightarrow t):=s\equiv v_F^0, v_F^1,\cdot \cdot \cdot,v_F^l,\cdot \cdot \cdot,v_F^{d-1},v_F^d\equiv t,
\end{equation}
where $v_F^l$ is the $l$-st stop and node $v_F^l$ belongs to layer $F\in
\{{A,B}\}$, and $d$ is the
number of stops in this path. Similar to Ref.~\cite{yan2006efficient}, we
denote an `efficient path' for any path between nodes $s$ and
$t$ as
\begin{equation}\label{eq2}
L(p(s \rightarrow t),\alpha_F,\beta_F)=
\sum_{l=0}^{n-1}\alpha_F[k(v_F^l)]^{\beta_F}.
\end{equation}
where $k(v_F^l)$ is the degree of node $v_F^l$
in layer $F$. The efficient path between $s$ and $t$ is corresponding
to the route that makes the sum $L(p(s \rightarrow t),
\alpha_F,\beta_F)$ minimum. If there are several efficient
paths between two nodes, we choose one randomly. The efficient path is related
to the macro-parameter $\alpha_F$ and micro-parameter
$\beta_F$. The macro-parameter $\alpha_F \geq 0$ controls
packet transmission speed in layer $F$, and reflects the macro-level transmission speed difference between layers. The smaller value of
$\alpha_F$, the faster transmission in layer $F$.
The parameter $\alpha_A$ ($\alpha_B$) corresponds to the slower (faster)
network and the ratio $\alpha_B/\alpha_A$ controls the relative time
spends each jump in layer $B$
compared with layer $A$. The micro-parameter
$\beta_F$ determines the tendency of packets' favour to
small-degree or large-degree nodes in layer $F$, and reflects
the micro-level difference between nodes in the same layer. Large degree
(small degree) nodes in layer $F$ are preferentially to be
the next stop when $\beta_F<0$ ($\beta_F>0$). When $\beta_F=0$,
nodes with different degrees have the same probability to be
the next stop. Fig.~\ref{fig1} illustrates the routing
on multilayer networks.

\section{theoretical analysis}

From the perspective of statistical physics, we use the order
parameter $H(R)$ to characterize the congestion on multilayer
networks~\cite{yan2006efficient},
\begin{equation}\label{eq3}
H(R)=\lim_{t\rightarrow\infty}\frac{C}{R}\frac{\langle\Delta W\rangle}{\Delta t},
\end{equation}
where $\Delta W=W(t+\Delta t)-W(t)$, $\langle\cdots\rangle$
is average value over $\Delta t$, and $W(t)$ is the total number
of accumulated packets in the system at time $t$.
From the varying of $H$ with $R$, we will
know the critical point $R_c$ (to be computed later) above
which the congestion occurs. For a small value of $R$, the number
of generated and delivered packets are balanced, i.e., every packet can
be transported to their destinations, thus $H(R)=0$. For a large
value of $R$ (i.e., $R> R_c$), the congestion occurs and the number
of accumulated packets increases with time, so $H(R)>0$. The
critical traffic capacity $R_c$ is the most significant parameter
of a transportation network, which can be used to evaluate the
performance of a routing strategy, i.e., the larger, the better.

To compute the value of $R_c$, we first define the efficient
betweenness centralities (EBC) of nodes in multilayer networks
as
\begin{equation}\label{eq4}
g(\alpha_F, \beta_F, v)=\sum_{s\neq t}\frac{\sigma^{st}
(\alpha_F, \beta_F, v)}{\sigma^{st}(\alpha_F, \beta_F)},
\end{equation}
where $\sigma^{st}(\alpha_F,\beta_F)$ is the number of
efficient paths between nodes $s$ and $t$ for given values of $\alpha_F$ and $\beta_F$, and $\sigma^{st}
(\alpha_F,\beta_F,v)$ is the number of efficient paths that pass
node $v$. The larger value of
$g(\alpha_F, \beta_F, v)$, the more efficient paths
that pass node $v$. As a result, node $v$ needs to process
more packets and has a larger probability to be congested.
We denote nodes with high values of EBC as high-load (HL) nodes, and similarly denote nodes
with low values of EBC as low-load (LL) nodes.
A coupled node $v^o$ can deliver the packets to its
neighbors in both layers $A$ and $B$, and
it has two values of EBCs $g(\alpha_F, \beta_F,
v_A^o)$ and $g(\alpha_F, \beta_F, v_B^o)$, where $v_A^o$ ($v_B^o$) represents
node $v^o$ in layer $A$ ($B$). If node $v_A^o$ or $v_B^o$ overload,
traffic congestion will occur at node $v^o$. Thus, node
$v^o$'s EBC in the system is the maximum value of $g(\alpha_F, \beta_F,
v_A^o)$ and $g(\alpha_F, \beta_F, v_B^o)$, i.e.,
$g(\alpha_F, \beta_F, v^o)=\max \{ g(\alpha_F, \beta_F, v_A^o), g(\alpha_F, \beta_F, v_B^o)\}$. For a non-coupled
node $v_A$ (i.e., node $v_A$ has no counterpart in layer $B$),
it can only deliver the packets to neighbors in layer $A$,
and its EBC can be expressed as $g(\alpha_F, \beta_F, v_A)$.

At every time step, the system will generate $R$ packets in
layer $A$. We can
get the average number of packets that a node $v$
needs to process as
\begin{equation}\label{eq5}
\langle \Theta_F\rangle=R\frac{g(\alpha_F, \beta_F, v)}{N_A(N_A-1)}.
\end{equation}
When $R\leq R_c$, there is no accumulated packets at any node in
the system, i.e., $\langle \Theta_F\rangle\leq C$.
When $R>R_c$, traffic congestion will occur at some HL nodes, i.e.,
$\langle \Theta_F\rangle>C$. Since the node with the largest EBC value has
the largest probability being congested, and combining
the condition $C=1$, the critical
packet generating number $R_c$ should fulfill
\begin{equation}\label{eq6}
R_c=\frac{N_A(N_A-1)}{g_{\rm max}(\alpha_F, \beta_F)},
\end{equation}
where $g_{\rm max}(\alpha_F,\beta_F)$ is the largest
value of EBC in the system for the given $\alpha_F$ and $\beta_F$.

\section{results}

We introduce four parameters to investigate the effectiveness
of CMR strategy. First, we introduce
a generalized parameter coupling based on
Ref.~\cite{morris2012transport}, which is used to describe how well two
layers are used to transmit the packets. Here the coupling is defined as
\begin{equation}\label{eq7}
\lambda = \frac{\sum\limits_{s\neq t}\sigma_{B}^{st}(\alpha_F, \beta_F)}{\sum\limits_{s\neq t}\sigma^{st}(\alpha_F, \beta_F)},
\end{equation}
where $\sigma^{st}(\alpha_F, \beta_F)$ is the number of efficient
paths between nodes $s$ and $t$ for given values of $\alpha_F$
and $\beta_F$, and $\sigma_{B}^{st}(\alpha_F, \beta_F)$ is
the number of efficient paths that contains
at least one edge in layer $B$. Specifically, we have
$\sigma_{B}^{st}(\alpha_F, \beta_F)=0$ when every efficient path between nodes $s$ and $t$ only uses the edges in layer $A$. For the
case of $\lambda\approx0$, most packets are transported only by
layer $A$, without using the edges in layer $B$. With the increase of $\lambda$, more packets are transported by using the edges
in $B$.

Secondly, we define $\delta$ as
\begin{equation}\label{eq8}
\delta = \frac{\sum\limits_{s\neq t}e_B^{st}(\alpha_F, \beta_F)}{\sum\limits_{s\neq t}e_A^{st}(\alpha_F, \beta_F)},
\end{equation}
where $e_A^{st}(\alpha_F, \beta_F)$ [$e_B^{st}(\alpha_F, \beta_F)$] is
the number of edges belonging to layer $A$ ($B$) in the efficient paths
between nodes $s$ and $t$ for given values of $\alpha_F$ and $\beta_F$.
When $\delta\approx0$, most edges that are used to deliver packets belong
to layer $A$. The more edges in layer $B$ are used to transport packets,
the larger value of $\delta$. The definitions of $\lambda$ and $\delta$
look similar, the difference is that coupling $\lambda$ represent the
proportion of all the efficient paths in system that contain edges
in layer $B$, while $\delta$ is that, in all efficient paths, the ratio of edges in $B$ and $A$.

Thirdly, we define the average length of efficient paths to
capture the effectiveness of the CMR strategy as
\begin{equation}\label{eq9}
\langle d \rangle= \frac{1}{N_A(N_A-1)}\sum\limits_{s\neq t} d^{st}(\alpha_F, \beta_F),
\end{equation}
where $d^{st}(\alpha_F, \beta_F)$ is the length or jumps of efficient paths
between node $s$ and $t$ for given values of $\alpha_F$ and $\beta_F$.
For example, the length of selected path between nodes $s$ and $t$ in
Fig.~\ref{fig1} is 3. The smaller of $\langle d \rangle$, the less average jumps of the packets arrive the destination.

To improve network traffic capacity, the
average packet delivery time $\langle T \rangle $ must be minimized. The
definition of $\langle T \rangle $ is
\begin{equation}\label{eq10}
\langle T \rangle=\lim_{t\rightarrow\infty} \frac{1}{n}\sum_{p=1}^{n}T_p,
\end{equation}
where $n$ is the number of arrived packets at a given time and $T_p$ is
the packet delivery time of packet $p$. The delivery time of each packet
consists of the travelling time from the origin to the destination and the
waiting time in the queue of the congested nodes. When $R$ is less
than $R_c$, $\langle T \rangle $ only depends on the travelling
time which is relatively small, while when $R>R_c$, $\langle T
\rangle $ increases with $R$ rapidly.

\subsection{Artificial multilayer networks}
In this subsection, we perform extensive numerical simulations
on artificial multilayer networks. We set the size of layer
$A$ as $N_A=1000$, degree exponents $\gamma=3.0$, the minimum degree
$k_{\rm min}=2$, and the maximum degree $k_{\rm max}\sim \sqrt{N_A}$. The size
of layer $B$ is $N_B=500$ and average degree $\langle k_B\rangle
=6$. All the results are obtained by averaging over 20
different network realizations, with 100 independent runs
on each realization.

\begin{figure}[h]
\centerline{\includegraphics[width=1\linewidth]{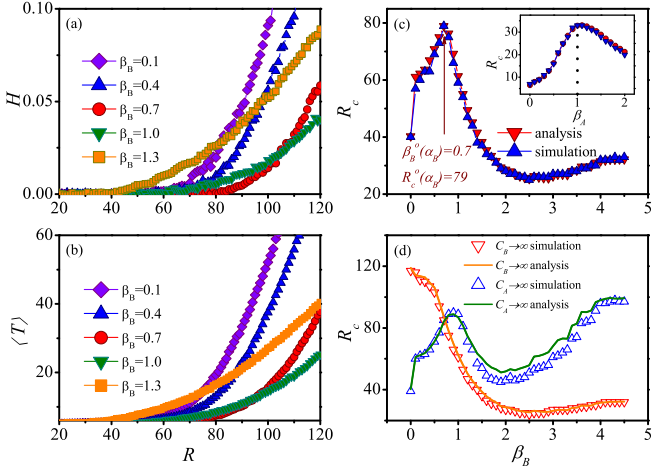}}
\caption{The efficiency of CMR strategy on artificial multilayer networks when $\alpha_B=0.5$. (a) The order parameter $H$ and (b) the average delivery
time $\langle T \rangle$ of packets versus $R$. (c) The traffic capacity $R_c$ of multilayer
networks and (d) the upper limit capacity of layer $B$ $(A)$ as a
function of the $\beta_B$. The inset of (c) is the traffic capacity
$R_c$ of monolayer networks (layer $A$) as a function of $\beta_A$ when $\alpha_A=1$. The theoretical analysis results are obtained from Eq.~(\ref{eq6}). We set other parameters as $N_A=1000$, $\gamma=3.0$,
$k_{\rm min}=2$, $k_{\rm max}\sim \sqrt{N_A}$, $N_B=500$, $\langle k_B\rangle =6$, $\alpha_A=1$, and $\beta_A=1$.}\label{fig2}
\end{figure}

\begin{figure}[htbp]
\centerline{\includegraphics[width=1\linewidth]{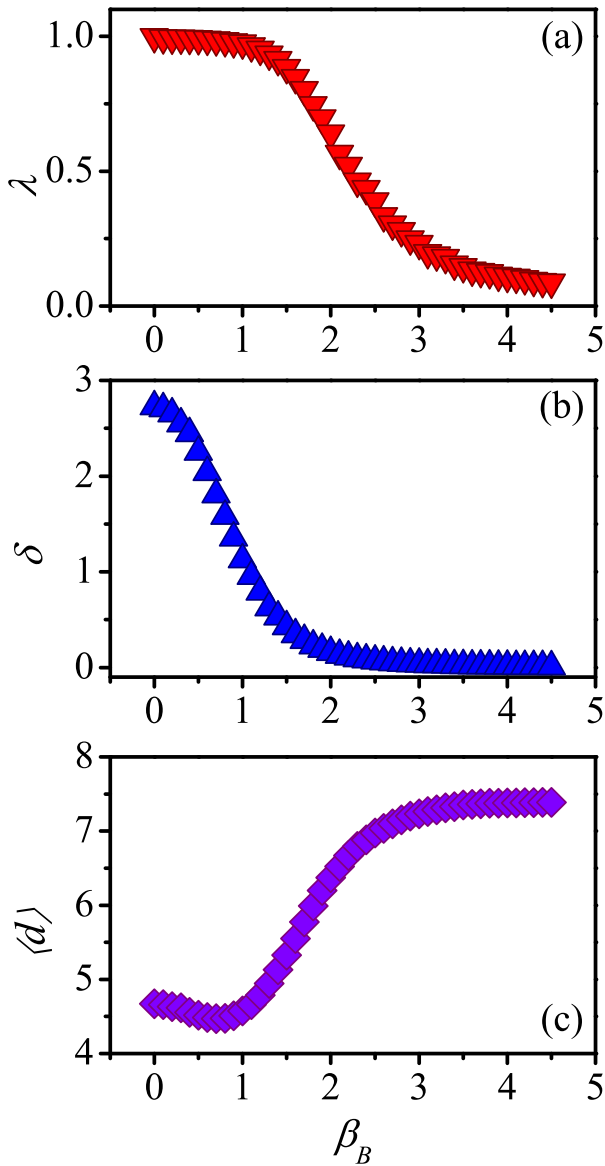}}
\caption{The parameters (a) coupling $\lambda$, (b) ratio of edges
that used in layers $B$ and $A$ $\delta$,
and (c) average jumps $\langle d \rangle$ between nodes VS $\beta_B$ on
artificial multilayer networks. We set other parameters as $N_A=1000$,
$\gamma=3.0$, $k_{\rm min}=2$, $k_{\rm max}\sim \sqrt{N_A}$, $N_B=500$,
$\langle k_B\rangle =6$, $\alpha_A=1$, $\beta_A=1$, and $\alpha_B=0.5$.}\label{fig3}
\end{figure}

We first focus on the effects of micro-parameter $\beta_B$ on the
effectiveness of CMR strategy in Fig.~\ref{fig2}. Since $\beta_A=1$
is optimal value without layer $B$ for the case of $\alpha_A=1$ [see the
inset of Fig.\ref{fig2}(c)], we set $\alpha_A=1$ and $\beta_A=1$.
Through extensive numerical simulations, we find that other values of
$\alpha_A $ and $\beta_A $ do not qualitatively affect the effectiveness
of the proposed CMR strategy. We set $\alpha_B/\alpha_A<1$ (i.e., $\alpha_B
\leq 1$) here, which indicates that a journey on the high speed layer
$B$ is favored for a journey in layer $A$. From Figs.~\ref{fig2}(a) and (b), we
find that for different values of micro-parameter $\beta_B$, both
the order parameter $H$ and average packet delivery time $\langle T
\rangle$ monotonically increases with $R$. Above the
threshold $R_c$, $H$ and $\langle T \rangle$ are finite, and increase
with $R$. Importantly, we find that $R_c$ exhibits a
non-monotonously varying with $\beta_B$ as shown in Fig.~\ref{fig2}(c),
and the system exists an optimal value $\beta_B^o(\alpha_B)=0.7$
at which the traffic capacity $R_c$ reaches the maximum value
$R_c^o(\alpha_B)=79$ when $\alpha_B=0.5$. The average length
of efficient paths $\langle d \rangle$ reaches the minimum
value at the same parameters [see Fig.~\ref{fig3}(c)]. Specifically,
$R_c$ first increases with $\beta_B$, and peaks at $\beta_B^o(\alpha_B)=0.7$,
and then decreases. The theoretical predictions agree well with the
numerical values of $R_c$. To understand the
non-monotonous phenomenon, we need to check what happens when
varying $\beta_B$. When $\beta_B$ is small (large), the values
of $\lambda $ and $\delta $ are large (small) as shown in
Figs.~\ref{fig3}(a) and (b). This indicates that packets
are more likely to be transmitted in layer $B$ ($A$).
For a small value of $\beta_B$,
many coupled nodes are used to transmit packets. Similar to
the effective strategy on monolayer networks, preferentially transmitting
the packets through small degree nodes in layer $B$ could
improve the traffic capacity of the system~\cite{yan2006efficient},
and $R_c$ thus first increases with $\beta_B$. For a large value of $\beta_B$, most packets
are transmitted on layer $A$, which decreases the usage of coupled nodes in
transmitting the packets, and $R_c$ thus decreases.
In Fig.~\ref{fig2}(d), we further verify in which layer
the congestion occurs for different values of $\beta_B$. To this
end, we set the delivery ability of nodes in layer $A$ ($B$) is
infinite when we check the upper limit capacity of layer
$B$ ($A$), i.e., $C_A\rightarrow\infty$ and $g(\alpha_F, \beta_F, v^o)=g(\alpha_F,
\beta_F, v_B^o)$ [$C_B\rightarrow\infty$ and $g(\alpha_F, \beta_F, v^o)=g(\alpha_F,
\beta_F, v_A^o)$]. We find that congestion occurs in layer $B$ ($A$)
for small (large) values of $\beta_B$, since layer $B$ ($A$) has
a smaller critical network throughput.
From what we discussed above, we can see that compared to the
isolated low speed network $A$, the capacity $R_c$ of
multilayer network is remarkably improved at some parameters,
since the traffic load of the low speed layer $A$ is
redistributed to the high speed layer $B$ reasonably.
The system capacity is affected by both layers $A$ and $B$, and depends
non-monotonically on micro-parameter $\beta_B$. The theoretical
predictions agree well with the numerical simulations in both
Figs.~\ref{fig2}(c) and (d).

\begin{figure}[htbp]
\centerline{\includegraphics[width=1\linewidth]{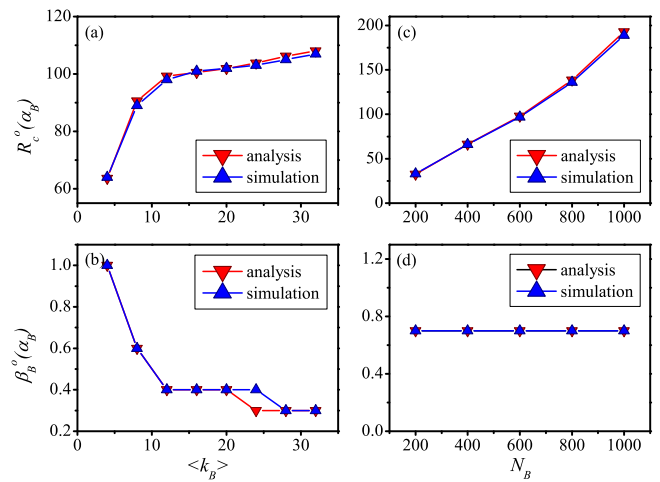}}
\caption{CMR strategy on artificial multilayer networks with different $\langle k_B\rangle$ and $N_B$. The maximum capacity $R_c^o(\alpha_B)$ (top), and the corresponding optimal micro-parameter $\beta_B^o(\alpha_B)$ (bottom) as a function of $\langle k_B\rangle$ [(a) and (b)] and $N_B$ [(c) and (d)]. We set other parameters as $N_A=1000$, $\gamma=3.0$, $k_{\rm min}=2$, $k_{\rm max}\sim \sqrt{N_A}$, $\alpha_A=1$, $\beta_A=1$, $\alpha_B=0.5$, $N_B=500$ [(a) and (b)], $\langle k_B\rangle =6$ [(c) and (d)].}\label{fig4}
\end{figure}

We further study the effects of network size $N_B$
and average degree $\langle k_B\rangle$ of layer $B$
(i.e., increasing the number of coupled nodes and edges
in the high speed network) on system capacity in
Fig.~\ref{fig4}. As shown in Figs.~\ref{fig4}(a) and (c), we find that
the maximum traffic capacity $R_c^o(\alpha_B)$ when $\alpha_B=0.5$ increases
with $\langle k_B\rangle$ and $N_B$, since the number of efficient paths
(coupled nodes) increase. That's to say, increasing the size and average degree of the high speed layer enhance the traffic capacity of multilayer networks effectively.
We note that the optimal micro-parameter $\beta_B^o(\alpha_B)$ decreases
with $\langle k_B\rangle$ [see Fig.~\ref{fig4}(b)], but does not
change with $N_B$ [see Fig.~\ref{fig4}(d)]. Again, our theoretical
predictions agree well with the numerical simulations.

\begin{figure}[htbp]
\centerline{\includegraphics[width=1\linewidth]{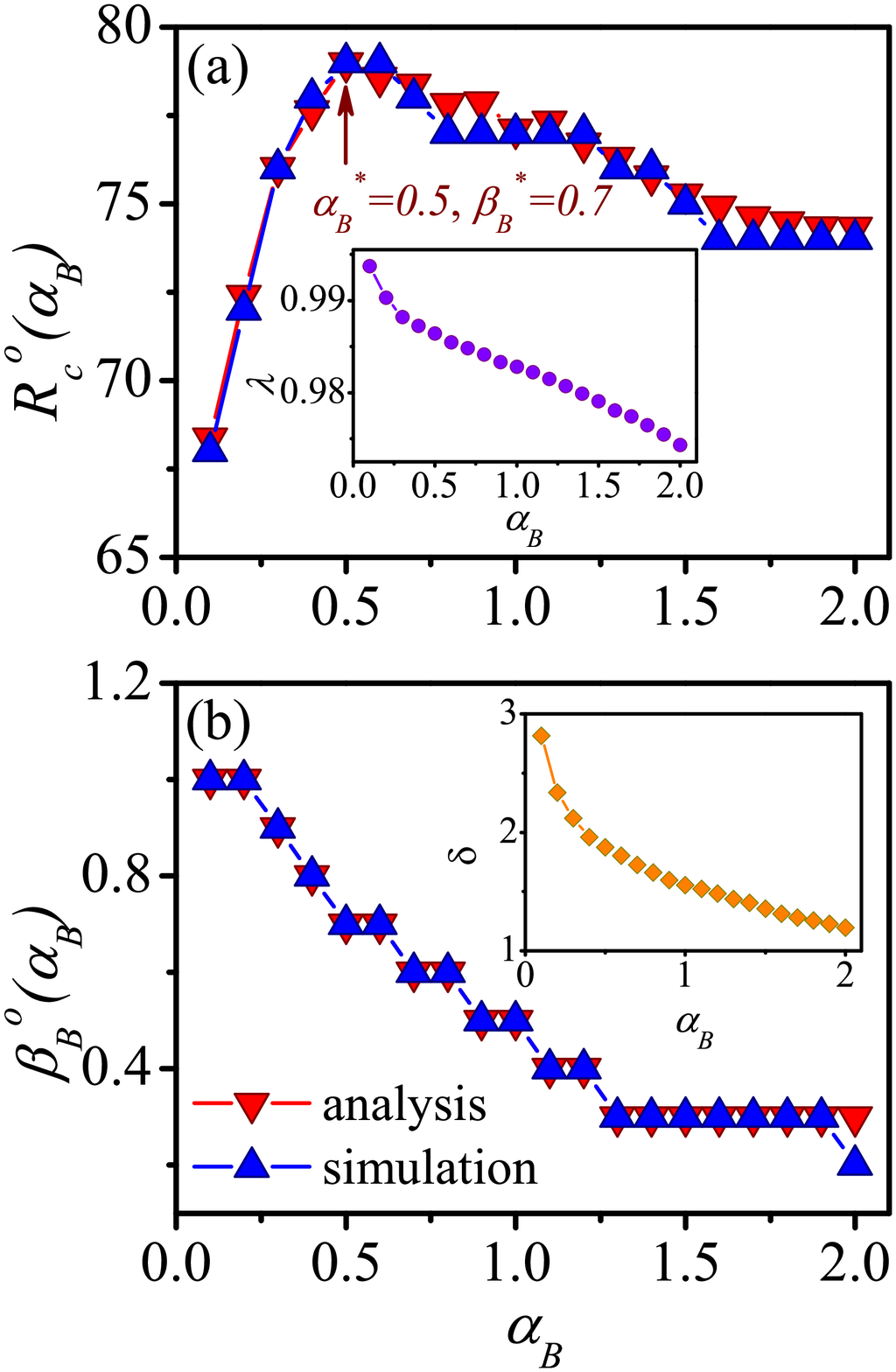}}
\caption{CMR strategy on artificial multilayer networks with different $\alpha_B$. (a) $R_c^o(\alpha_B)$ and (b) $\beta_B^o(\alpha_B)$ as a function of $\alpha_B$. The insets of (a) and (b)
respectively exhibit $\lambda$ and $\delta$ versus $\alpha_B$ when $\beta_B=\beta_B^o(\alpha_B)$.
We set other parameters as $N_A=1000$, $\gamma=3.0$, $k_{\rm min}=2$, $k_{\rm max}\sim \sqrt{N_A}$, $N_B=500$, $\langle k_B\rangle =6$, $\alpha_A=1$, $\beta_A=1$.}\label{fig5}
\end{figure}

All results above are obtained when macro-parameter $\alpha_B=0.5$, we
next study the effects of macro-parameter $\alpha_B$ in Fig.~\ref{fig5}.
We find that $R_c^o(\alpha_B)$ depends non-monotonically on $\alpha_B$
[$R_c^o(\alpha_B)$ first increase with $\alpha_B$ and then decrease], and
the corresponding optimal micro-parameter $\beta_B^o(\alpha_B)$ monotonically
decreases with $\alpha_B$ and reaches a suitable
packets' preference to layer $B$. For small (large) values of $\alpha_B$,
$\lambda $ and $\delta $ are large (small) as shown in the insets of
Figs.~\ref{fig5}(a) and (b) respectively. Importantly, we find that
the system reaches a maximum traffic capacity $R_c^\star$ at the optimal
macro- and micro-level parameters combination $(\alpha_B^\star, \beta_B^\star)$,
and the number of delivered packets by each layer reach to a balance. From
Figs.~\ref{fig5}(a) and (b), we obtain $R_c^\star=79$ at $(\alpha_B^\star,
\beta_B^\star)=(0.5,0.7)$. Without the high speed network $B$, the
maximum traffic capacity of low speed network is 33 [see the inset of
Fig.~\ref{fig2}(b)]. The maximum traffic capacity of system is improved
about 2.5 times once the network $B$ is induced. Although establishing
high speed transportation can improve the traffic capacity of low
speed network, our results indicate that a reasonable redistribution
of traffic load is an essential issue. The theoretical predictions
agree well with the numerical simulations.

\subsection{Real-world networks}
A wide range of systems in the real world have multiple
subsystems and layers of connectivity, which can be described
as multilayer networks~\cite{gao2012networks,kivela2014multilayer,lee2015towards,
boccaletti2014structure}. We verify the effectiveness
of our proposed CMR strategy on a real-world multilayer network, which
is a social network of Employees of Computer Science Department
(ECSD) at Aarhus University~\cite{magnani2013combinatorial}. The
multilayer social network consists of five kinds of online and
offline relationships (Facebook, Leisure, Work, Co-authorship, Lunch),
and we choose the Work and Facebook relationships as layer $A$ and
layer $B$, respectively. We denote this real-world network
as Work-Facebook multilayer network. Layer $A$ composes of $60$ nodes
and $194$ edges, and layer $B$ has $32$ nodes and $124$ edges.
Some structural properties of the two networks are presented in Table~\ref{table1}.

\begin{figure}
\begin{center}
\epsfig{file=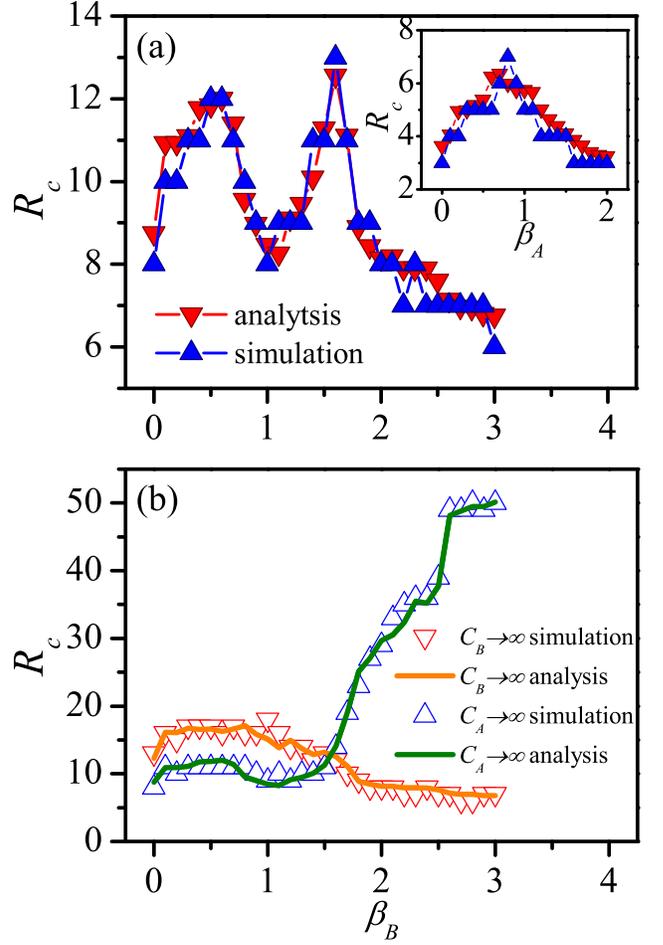,width=1\linewidth}
\caption{CMR strategy on Work-Facebook multilayer network with $\alpha_B=0.5$. (a) The traffic capacity $R_c$, and (b) the upper limit capacity of layer $A$ (Work) and layer $B$ (Facebook) as a function of the $\beta_B$. The inset of (a) is the traffic capacity $R_c$ of isolated Work network as a function of $\beta_A$ when $\alpha_A=1$. We set $\alpha_A=1$, $\beta_A=1$, and $\alpha_B=0.5$.}
\label{fig6}
\end{center}
\end{figure}

\begin{figure}[htbp]
\centerline{\includegraphics[width=1\linewidth]{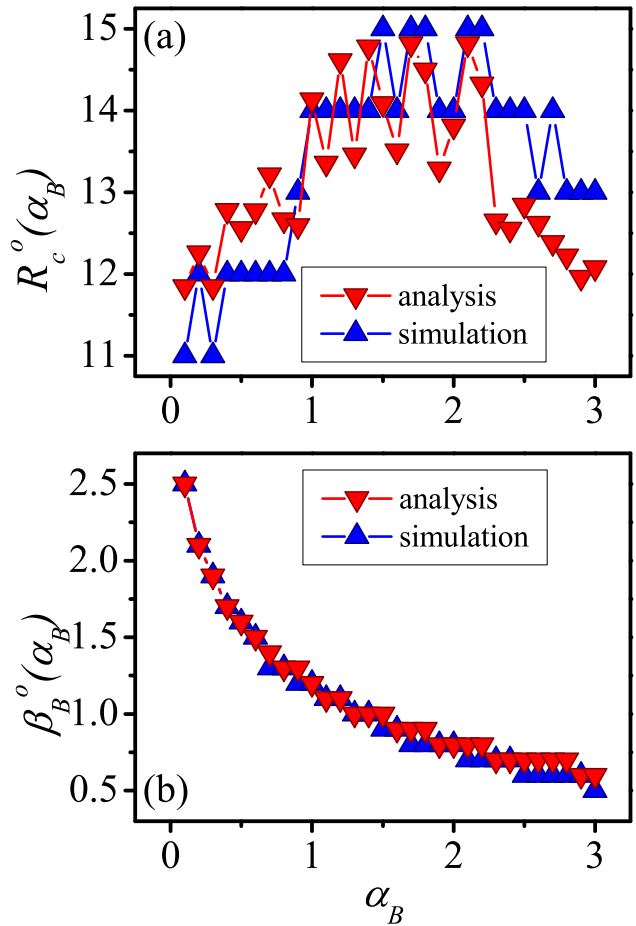}}
\caption{CMR strategy on Work-Facebook multilayer network with different $\alpha_B$. (a) The maximum capacity $R_c^o(\alpha_B)$ for a given $\alpha_B$, and (b) the corresponding optimal micro-parameter $\beta_B^o(\alpha_B)$ as a function of macro-parameter $\alpha_B$. We set $\alpha_A=1$, $\beta_A=1$.}\label{fig7}
\end{figure}

\begin{table}
  \caption{\label{table1} Structural properties of Work network and Facebook network, including number of nodes $N$, number of edges $E$, mean degree $\langle k \rangle$, maximum degree $k_{\mathrm{max}}$, degree heterogeneity $H_k=\langle k^2 \rangle/\langle k \rangle^2$, diameter $D$, average shortest distance $ L $, correlation coefficient $r$, clustering coefficient $c$, and modularity $Q$. }
\begin{tabular}{c c c c c c c c c c c c }
\hline
Network & $N$ & $E$ & $\langle k \rangle$ & $ k_{\mathrm{max}} $& $ H_{k} $ & $D$ &$  L $& $r$ & $c$ & $Q$ \\
\hline
Work & $60$ & $194$ & $6.5$ & $27$ & $1.7$ & $4$ & $2.4$ & $-0.218$ & $0.64$ & $0.46$\\
Facebook & $32$ & $124$ & $7.8$ & $15$ & $2.3$ & $4$ & $2.0$ & $0.003$ & $0.54$ & $0.34$\\
\hline
\end{tabular}
\end{table}

We study the effectiveness of the CMR strategy on the
Work-Facebook multilayer network in Fig.~\ref{fig6}.
Since the extremely complicated structures of networks,
there has two peaks of $R_c$ versus $\beta_B$, at which
the traffic capacity is very large, and the $R_c^o(\alpha_B)$
corresponds to the second peak. Compared to the
isolated Work network, the capacity $R_c$ of
Work-Facebook multilayer network are improved
when Facebook network joins in the system [see
the inset of Fig.~\ref{fig6}(a)], and the system
capacity is affected by both Work and Facebook networks
[see Fig.~\ref{fig6}(b)].
In Fig.~\ref{fig7}, we find that $R_c^o(\alpha_B)$ versus $\alpha_B$
exhibits a nonmonotonic pattern [see Fig.~\ref{fig7}(a)], and
the corresponding optimal micro-parameter $\beta_B^o(\alpha_B)$ monotonically
decreases with $\alpha_B$ [see Fig.~\ref{fig7}(b)].
Similar to the artificial networks, we find that
the system reaches a maximum traffic capacity $R_c^\star=15$
at the optimal micro- and macro-level parameters
$(\alpha_B^\star, \beta_B^\star)=(2.1,0.7)$.
The fluctuation of curves in Figs.~\ref{fig6} and~\ref{fig7}
is caused by the extremely complicated structures of both
Work and Facebook networks. We should note that the theoretical
predictions markedly well agree with the numerical simulations.

\section{DISCUSSIONS}

For the purpose of alleviating
the congestion of a low speed transportation network,
an intuitive way is to built a new high speed network in
busy regions or among the high flow nodes. The low
and high speed networks constitute a multilayer network.
How to reasonable redistribution of traffic load to
maximize the multilayer network is an essential issue
and full of challenges.
In this work, we first proposed a comprehensive multilayer
network routing (CMR) strategy by considering different transmission
speeds of layers from the macroscopic view (by adjusting a macro-parameter $\alpha_F$), and different roles
of nodes from the perspective of microscopic structure (controlled by a adjustable micro-parameter $\beta_F$). We then performed extensive
numerical simulations on both artificial and real-world
networks. We found that our routing strategy can redistribute
the traffic load in low speed layer to high
speed layer reasonably, and the traffic capacity of multilayer network are
remarkably enhanced compared with the monolayer low speed network. In addition,
the system capacity is affected by both layers $A$ and $B$, and depends
non-monotonically on micro-parameter $\beta_B$ and macro-parameter
$\alpha_B$. For a given multilayer network, the system reaches
a maximum traffic capacity $R_c^\star$ at the optimal
micro- and macro-level parameters $(\alpha_B^\star, \beta_B^\star)$.
Moreover, we found that increasing
the size and the average degree of the high speed layer $B$
enhances the transport capacity of multilayer networks more
effectively. The theoretical predictions agree well with the numerical
simulations in both artificial and real-world networks.

A wise way to alleviate traffic congestion for multilayer
networks is designing effective multilayer network routing
strategy. Our results exhibit
a way to reasonable redistribute the traffic load. In this
work, we proposed an effective strategy
which considers the local structures of different nodes, as well as
the transmission speeds of different layers. We study our
proposed strategy on multilayer networks including two layers,
and it can remarkably improve the systems' traffic capacity.
Our research may stimulate future studies
on designing realistic transportation and communication multilayer
networks, such as, considering
different delivery abilities of nodes,
limited traffic resources, transmission cost of layers,
and multilayer networks with more than two layers.

\acknowledgments
This work was supported by the National Natural Science Foundation of China (Grant Nos. 11575041 and 61673086), and the Fundamental Research Funds for the Central Universities (Grant No. ZYGX2015J153).


\end{document}